\documentclass[a4paper]{jpconf}
\usepackage[english]{babel}
\usepackage{graphicx}
\usepackage{mathtools}
\usepackage{color}
\newcommand*{\I}{\mathrm{i}}

\begin{document}
\title{Low energy quantum regimes of 1D dipolar Hubbard model with correlated hopping}
\author{S Fazzini$^1$, L Barbiero$^{2,3}$ and A Montorsi$^1$} 
\address{$^1$ Institute for condensed matter physics and complex systems, DISAT, Politecnico di Torino, I-10129, Italy}
\address{$^2$ CNR-IOM DEMOCRITOS Simulation Center and SISSA, Via Bonomea 265, I-34136 Trieste, Italy}
\address{$^3$ Dipartimento di Fisica e Astronomia ``Galileo Galilei", Universit\`a di Padova, 35131 Padova, Italy}

\ead{serena.fazzini@polito.it}

\begin{abstract} We apply the bosonization technique to derive the phase diagram of a balanced unit density two-component dipolar Fermi gas in a one dimensional lattice geometry. The considered interaction processes are of the usual contact and dipolar long-range density-density type together with peculiar correlated hopping terms which can be generated dynamically. Rigorous bounds for the transition lines are obtained in the weak coupling regime.  In addition to the standard bosonization description, we derive the low energy phase diagram taking place when part of the interaction is embodied non-perturbatively in the single component Hamiltonians.  In this case the Luttinger liquid regime is shown to become unstable with respect to the opening of further gapped phases,  among which insulating bond ordered wave and Haldane phases, the latter with degenerate edge modes. 
\end{abstract}
\section{Introduction}
Experiments with cold atoms have disclosed a new way of investigating strongly correlated systems \cite{Bloch2008}. The possibility to cool down both fermionic and bosonic atomic gases to very low temperature and to trap them onto optical lattices, together with the ability to tune the interactions and the dimensionality with high accuracy, has allowed to simulate a great variety of interacting many-body lattice Hamiltonians \cite{dutta}. Particular theoretical efforts have been devoted to identifying fermionic Hubbard-like systems \cite{dutta1,BMR,FMRB} which ground state exhibits exotic \cite{Haldane} and topological phases \cite{Wen}, described by string-like order parameters \cite{dennijs}. Also particles with dipolar long range interaction like polar molecules and magnetic atoms \cite{review_Santos} are currently available in laboratories. This has allowed for the experimental realization of a paradigmatic model in condensed matter physics namely the extended Bose-Hubbard model \cite{Ferlaino_Zoeller}.\\
On the other hand, bosonization \cite{giam, nersesian} is a well-established analytical technique to investigate the low energy regime of one dimensional interacting fermionic systems. It consists in linearizing the spectrum around the Fermi points, passing to the continuum limit, and finally expressing the fermionic operators in terms of bosonic fields. In this way one has an efficient and general setting to study the low energy excitations induced by different types of interaction. Despite the fact that in most cases further numerical analyses are needed in order to get the full zero temperature phase diagram, bosonization is very useful to understand the nature of many quantum phases of matter and evaluate the correlation functions which characterize them.
\\Within bosonization approach, usually the interaction is included in a perturbative way, starting from two non-interacting single component Luttinger liquids (LLs). In fact, as noticed in \cite{giam}, in some cases part of the interaction can be included non-perturbatively already in the single component Hamiltonian, as long as it remains in a LL regime. This possibility has been exploited for instance in \cite{DBRD} to predict the presence of a bond ordered wave (BOW) phase induced by dipolar interaction already within one-loop bosonization. Here we adopt it to derive the zero temperature phase diagram of a one dimensional Hubbard model in presence of correlated hopping -- induced by a periodical modulation of the on-site interaction \cite{diliberto}, as recently shown experimentally \cite{meinert}-- and long-range dipole-dipole interaction.\\
The paper is organized as follows. In section 2 we introduce the model and rewrite the Hamiltonian in  normal ordered form. In section 3 we review its bosonization phase diagram, which coincides with that of two spin-charge decoupled sine- Gordon models, due to the peculiar $r^{-3}$ nature of the power law decay of the dipolar interaction. The different gapped and partly gapped phases are characterized in terms of string and parity \cite{BDGA, MR} non-local orders.  In section 4 we then revisit the phase diagram by including part of both dipolar and correlated hopping interaction non perturbatively in the single component Hamiltonians. In this case, the phase diagram obtained from the study of renormalization group (RG) flow equations exhibit a richer structure. In section 5 we give some conclusions.
\section{The model}
We consider a  balanced unit density two-component dipolar Fermi mixture. Once these particles are trapped in a one dimensional lattice, an accurate description of the system is given  by the following Hamiltonian \cite{FMRB}
\begin{equation}
H=-J\sum_{j,\sigma}Q_{j,j+1, \sigma}\left[1-X(n_{j,\bar{\sigma}}-n_{j+1,\bar{\sigma}})^2\right]
+U_0\sum_j n_{j,\uparrow}n_{j,\downarrow}+V\sum_{j,r>0}\frac{ n_j n_{j+r}}{r^3}
\end{equation}
where $Q_{j,j+1,\sigma}=c_{j,\sigma}^{\dagger}c_{j+1,\sigma} + h.c.$;  $\sigma=\uparrow,\downarrow$ is the index species and $\bar{\sigma}$ denotes its opposite, $c_{j,\sigma}^{\dagger}$ and $c_{j,\sigma}$ are the creation and annihilation operators, respectively, $n_{j,\sigma}$ counts the number of particles of species $\sigma$ and $n_j=\sum_{\sigma}n_{j,\sigma}$. The coupling coefficients $J,U_0,V,X$, independently tunable in the experiments \cite{Bloch2008, Goral2003, bartolo}, describe the tunneling probability, on-site and dipolar interactions, and correlated hopping processes, respectively. Upon normal ordering of the operators, $Q_{j,j+1,\sigma}=:Q_{j,j+1,\sigma}:+\langle Q_{j,j+1,\sigma}\rangle$, $n_{j,\sigma}=:n_{j,\sigma}:+\langle n_{j,\sigma}\rangle$ (with $\langle Q_{j,j+1,\sigma}\rangle=2/\pi$, $\langle n_{j,\sigma}\rangle=1/2$),
and omitting the constant terms, we get
\begin{equation}
\label{HamNormOrd}
\begin{split}
& H=-\left(1-\frac{X}{2}\right)\sum_{j,\sigma}:Q_{j,j+1,\sigma}:+\sum_{j,\sigma}\sum_rV_{\parallel}(r):n_{j,\sigma}::n_{j+r,\sigma}:+\\
&+\sum_{j,\sigma}\sum_rV_{\perp}(r):n_{j,\sigma}::n_{j+r,\bar{\sigma}}:+U_0\sum_j:n_{j,\uparrow}::n_{j,\downarrow}:\\&
-2 X \sum_{j,\sigma}:Q_{j,j+1,\sigma}::n_{j,\bar{\sigma}}::n_{j+1,\bar{\sigma}}:
\end{split}
\end{equation}
where we have set $J=1$ and have defined
$V_{\parallel}(r)=\frac{V}{r^3}-\frac{4X}{\pi}\delta_{r,1}$,
$V_{\perp}=\frac{V}{r^3}$.

\section{Weak coupling phase diagram}
In the standard bosonization approach, one starts from the non-interacting Hamiltonian and consider the effect of interactions in a perturbative manner. The first step is to perform the continuum limit:
$
\sum_j\longrightarrow \frac{1}{a}\int dx $;
$c_{j,\sigma}\longrightarrow\sqrt{a}\left[e^{\I k_Fx}\Psi_{R\sigma}(x)+e^{-\I k_Fx}\Psi_{L\sigma}(x)\right]
$
(with $x=ja$, $a\to 0$ being the lattice constant). Here $\Psi_{R\sigma}$ and $\Psi_{L\sigma}$ are the fermionic field operators for the right and left movers, respectively. As claimed before, we will consider the particular case in which the system is at half filling; hence $k_F=\pi/(2a)$.
We finally write the fermionic fields $\Psi_{\chi\sigma}$ in terms of the bosonic ones $\phi_{\sigma}$ and $\theta_{\sigma}$:
\begin{equation}
\label{Campo_fermionico}
\Psi_{\chi\sigma}(x)=\frac{\eta_{\chi\sigma}}{\sqrt{2\pi\alpha}}e^{\I\sqrt{\pi}\left[\chi\phi_{\sigma}(x)+\theta_{\sigma}(x)\right]}
\end{equation}
where the generic index $\chi$ can denote right (R) or left (L) movers, with the usual convention that it assumes a positive sign in the first case and a negative sign in the second one. Here $\alpha\sim a$ is an ultraviolet cutoff and $\eta_{\chi\sigma}$ are the Klein factors, which guarantee proper anti-commutation relations. With this notation, the kinetic operator $:Q_{j,j+1,\sigma}:$ and the density operator $:n_{j,\sigma}:$ appearing in Hamiltonian (\ref{HamNormOrd}) have the following expressions in terms of the bosonic fields
\begin{equation}
\label{:Q:}
:Q_{j,j+1,\sigma}:= -\frac{2}{\pi}(-1)^j\cos{(2\sqrt{\pi}\phi_{\sigma}(x))}-a^2\left[\left(\nabla\phi_{\sigma}(x)\right)^2+\left(\nabla\theta_{\sigma}(x)\right)^2\right]+...
\end{equation}
\begin{equation}
\label{:n:}
:n_{j,\sigma}:=a\left[\frac{1}{\sqrt{\pi}}\nabla\phi_{\sigma}(x)-\frac{(-1)^j}{\pi a}\sin{(2\sqrt{\pi}\phi_{\sigma}(x))}\right]+...
\end{equation}
\noindent Here we have used dots to denote the higher order terms in expansion with respect to $a$, which will be neglected.
Bosonization of two and three body-terms in the Hamiltonian (\ref{HamNormOrd}) involves calculating the product of operators of the form (\ref{:Q:}) and (\ref{:n:}). When the  latters act on different fermionic species, the calculation is straightforward. When acting on the same species instead, the operator product expansion is needed. In deriving the bosonized expression for $:n_{j,\sigma}::n_{j+r,\sigma}:$, we make use of the fusion rule 
$
 \sin{(2\sqrt{\pi}\phi_{\sigma}(x))}\sin{(2\sqrt{\pi}\phi_{\sigma}(x+R))}=\frac{a^2}{2R^2}-\frac{1}{2}\cos{(4\sqrt{\pi}\phi_{\sigma}(x))}-\pi a^2\left(\nabla\phi_{\sigma}(x)\right)^2+...
$,
 with $R=ra$.
  Thus we get
 \begin{equation}
 \label{:nn_r:}
 :n_{j,\sigma}::n_{j+r,\sigma}:\simeq  a^2\left\{  \frac{1-(-1)^r}{\pi}\left(\nabla\phi_{\sigma}(x)\right)^2-\frac{(-1)^r}{2\pi^2a^2}\cos{(4\sqrt{\pi}\phi_{\sigma}(x))}+\frac{(-1)^r}{2\pi^2R^2} + (-1)^j...\right\}. 
 \end{equation}
  In the three-body term, also the oscillating part of (\ref{:nn_r:}) (with $r=1$) contributes. To compute it, we apply the following operator product expansion,
$
  \nabla\phi_{\sigma}(x)\sin{(2\sqrt{\pi}\phi_{\sigma}(x+a))}=-\sin{(2\sqrt{\pi}\phi_{\sigma}(x))\nabla\phi_{\sigma}(x+a)}=\frac{1}{\sqrt{\pi}a}\cos{(2\sqrt{\pi}\phi_{\sigma}(x))}+...
$.
  One obtains $  (-1)^j\frac{2}{\pi^2a^2}\cos\left(2\sqrt{\pi}\phi_{\sigma}(x)\right)$, from which the three body term contribution is evaluated.
  As customary, it is now convenient to introduce charge and spin field operators, defined as
 $
  \phi_c(x)=\left(\phi_{\uparrow}(x)+\phi_{\downarrow}(x)\right)/\sqrt{2}$ and $\phi_s(x)=\left(\phi_{\uparrow}(x)-\phi_{\downarrow}(x)\right)/\sqrt{2}$, respectively.   
  Similar relations hold for the dual fields $\theta_{\nu}$ ($\nu=c,s$) as functions of $\theta_{\sigma}$ ($\sigma=\uparrow,\downarrow$). 
  In terms of these operators, the Hamiltonian can be separated into the sum of two independent Hamiltonians in the charge and spin sectors plus a coupling term:
  \begin{equation}
  H=H_c+H_s+H_{cs} \quad .
  \end{equation}
  In each sector $H_{\nu}$ has the the form of a sine-Gordon model:
  \begin{equation}
  \label{H_separate}
  H_{\nu}=\frac{v_{\nu}}{2}\int dx \left[\left(\sqrt{K_{\nu}}\nabla\theta_{\nu}\right)^2+\left(\frac{\nabla\phi_{\nu}}{\sqrt{K_{\nu}}}\right)^2\right]+\frac{m_{\nu}v_{\nu}}{2\pi a^2}\int dx \cos{(\sqrt{8\pi}\phi_{\nu})},\quad \nu=c,s;
  \end{equation}
  whereas the coupling Hamiltonian $H_{cs}$ reads
  $H_{cs}=\frac{M_{cs}}{\pi a}\int dx \cos{(\sqrt{8\pi}\phi_c)\cos{(\sqrt{8\pi}\phi_s)}}$.
   The coefficients of the Luttinger and mass terms are defined as follows:
   \begin{eqnarray}
   &K_{\nu}=1+\frac{1}{4\pi}\left[\frac{16X}{\pi}-s_{\nu}U_0-\frac{3\zeta(3)V}{2}-4\zeta(3)V\delta_{\nu,c}\right]\quad ;\quad v_{\nu}=2a\left[2-\frac{X}{2}-\frac{X}{\pi^2}-K_{\nu}\right]\\
   &m_{\nu}=\frac{1}{2\pi}\left[\frac{16X}{\pi}-s_{\nu}\left(U_0-\frac{3\zeta(3)V}{2}\right)\right]\quad ;\quad  
   M_{cs}=\frac{1}{2\pi}\left[\frac{3\zeta(3)V}{2}-\frac{8X}{\pi}\right].
   \end{eqnarray}
   where $s_c=1$, $s_s=-1$ and $\zeta(n)$ denotes the Riemann zeta function.
   The competition of the kinetic and mass terms in (\ref{H_separate}) can be discussed by analyzing the RG flow equations as in \cite{BMR}. Apart from the gapless LL phase, two insulating charge gapped phases are associated to the pinning of solely the charge field to the two values $\phi_c=0,\sqrt{\frac{\pi}{8}}$. In a uniform unit-density background, the latter values of $\phi_c$ describe respectively a Mott insulating (MI) regime with localized holon-doublon fluctuations and a Haldane insulator (HI) phase {\cite{dallatorre}} with ``dilute" hidden antiferromagnetic order of holons and doublons {in analogy to the $XXZ$ spin-1 chain behavior \cite{Haldane,dennijs}}. When $\phi_c$ instead is unpinned, a metallic spin gapped phase is associated to the pinning of solely the spin field to the value $\phi_s=0$. It actually describes the Luther Emery  (LE) \cite{LE} liquid phase where spin-up and spin-down single particle quantum fluctuations take place in a uniform background of condensed holons and doublons. The $\phi_s=0$ value also supports two insulating fully gapped regimes, occurring when the charge field is pinned as well.  In this case for $\phi_c=0$ the bond ordered wave (BOW) phase takes place. Whereas for $\phi_c=\sqrt{\frac{\pi}{8}}$ a charge density wave (CDW) phase with holon-doublon antiferromagnetic ordering is obtained. 
The five gapped or partly gapped phases can be characterized by the non-vanishing of the appropriate parity or string non-local order parameter, defined respectively as
\begin{equation}
O_P^\nu(j)=\langle\prod_{k=0}^{j-1}{\rm e}^{i\pi S_{z,k}^\nu}\rangle\sim  \langle\cos(\sqrt{2\pi}\phi_{\nu})\rangle \hspace{4pt} ,\hspace{4pt} O_S^\nu(j)=\langle\left (\prod_{k=0}^{j-1}{\rm e}^{i\pi S_{z,k}^\nu}\right )S_{z,j}^\nu\rangle\sim  \langle\sin(\sqrt{2\pi}\phi_{\nu})\rangle \hspace{4pt} .
\end{equation}
This is reported in Table \ref{table1}, following the procedure outlined in \cite{MR,BMR}. In particular, the HI phase turns out to have non-trivial topological properties, as the presence of degenerate edge modes \cite{MDIR}.
\begin{table}

	\vspace{-0.5cm}
	\caption{\label{table1}Correspondence between ground state quantum phases and nonlocal operators that manifest long range order (LRO) \cite{BMR}. We have indicated the unpinned fields with the letter \textit{u}.}
	\begin{center}
\begin{tabular}{llllll}
	\br 
 & $\sqrt{2\pi}\Phi_{c}$ & $\sqrt{2\pi}\Phi_{s}$ & $\Delta_{c}$ & $\Delta_{s}$ & LRO
 \\
\mr 
LL & $u$ & $u$ & 0 & 0 & none\\
LE & $u$ & 0 & 0 & open & $O_{P}^{s}$\\
MI & 0 & $u$ & open & 0 & $O_{P}^{c}$\\
HI & $\pi/2$ & $u$ & open & 0 & $O_{S}^{c}$\\
BOW & 0 & 0 & open & open & $O_{P}^{c},\: O_{P}^{s}$\\
CDW & $\pi/2$ & 0 & open & open & $O_{S}^{c},\, O_{P}^{s}$\\
\br
\end{tabular}
\end{center}
\vspace{-0.5cm}
\end{table}
Depending on the values of the coupling constants $J,U_0,V,X$ all the above regimes can be realized. For instance, as reported in Fig. 1, already at $V=0$ by varying $U_0$ and $X$ the LL, MI and LE regimes are achieved. 

   \section{Including interaction non-perturbatively}
\begin{figure}
	\vspace{-0.4cm}
	\centering
	\includegraphics[scale=0.3]{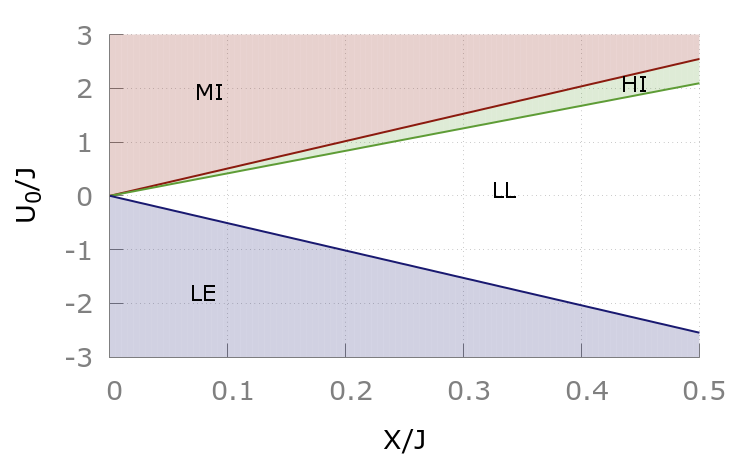}
	\caption{Phase diagram at $V=0$ from bosonization analysis of section 4. In case of standard bosonization of section 3, the Haldane insulator phase HI would still be a LL.}
	\label{fig1}
	\vspace{-0.2cm}
\end{figure}

   To gain further insight into the zero temperature phase diagram, we may regard Hamiltonian (\ref{HamNormOrd}) as the sum of two single-species Hamiltonians, already containing part of the interaction non perturbatively, plus an inter-species part: 
   \begin{equation}
   H=\sum_{\sigma}H_{\sigma}+H_{\uparrow\downarrow} \quad .
   \end{equation}
   Here $\sum_{\sigma}H_{\sigma}$ is given by the first two terms in (\ref{HamNormOrd}). Up to a multiplicative constant $(1-\frac{X}{2})$, the single-species Hamiltonian $H_{\sigma}$ is a long-range ``t-V" model, which is known to have a gapless Luttinger liquid phase for small enough interaction strength. In this case the Luttinger parameter $K(V,X)$ can be evaluated numerically with high precision \cite{CiOr, Ot}, and analytically both for vanishing $X$ \cite{PuZo}, and for vanishing $V$ \cite{giam}. Explicitly:
\begin{equation}
K(V,0)= \left[1+\frac{6 \zeta(3)V}{\pi^2}\right]^{-1/2} \quad , \quad
K(0,X)=\left[\frac{2}{\pi} \arccos\frac{2 X}{\pi\left(1-\frac{X}{2}\right)}\right]^{-1} \quad . 
\label{Kex}
\end{equation} 
Of course, $K$ can also be obtained within a bosonization approximation \cite{cap}. In this case
    $K(X,V)\simeq 1+\frac{1}{4\pi}\left[\frac{16X}{\pi}-\frac{7\zeta(3)V}{2}\right]$.
At this point one can proceed to bosonization of the inter-species interaction $H_{\uparrow\downarrow}$. It can be added to the Luttinger liquid phase of the remaining part of the Hamiltonian, which Luttinger coefficients may be determined either analytically or numerically \cite{CiOr, Ot}.  In so doing one ends up  with a Hamiltonian which is fully decoupled in the charge and spin fields. Explicitly:
   \begin{equation}
   H=H_c'+H_s'
   \end{equation}
   with
   \begin{equation}
   H_{\nu}'=\frac{v_{\nu}'}{2}\int dx \left[K_{\nu}'\left(\partial_x\theta_{\nu}\right)^2+\frac{1}{K_{\nu}'}\left(\partial_x\phi_{\nu}\right)^2\right]+\frac{m_{\nu}v_{\nu}}{2\pi a^2}\int dx \cos\left(\sqrt{8\pi}\phi_{\nu}\right) \label{Hamp}
   \end{equation}
   and
\begin{equation}
K_{\nu}'=\sqrt{K}\left\{1-\frac{X}{4\pi^2}+\frac{K^2}{8\pi}\left[\frac{2X}{\pi}-s_{\nu}\left(U_0+2\zeta(3)V\right)\right]\right\}\quad , \quad v_{\nu}'=v_{\nu}\frac{K_\nu}{K_{\nu}'}
\end{equation}
   where $s_c=1$, $s_s=-1$.
   Now the study of RG flow equations can be done by inserting the non-perturbative dependence of $K$ on $V$ ($X$), while keeping to first order the remaining interaction in $K$, $K_{\nu}'$ and $m_{\nu}$. The case $X=0$ has already been treated in this way in \cite{DBRD}. At variance with what obtained with standard bosonization, the latter approach allows to identify a further BOW phase already within one-loop expansion, thanks to the decoupling of the opening of the spin gap from the Gaussian transition of the charge gap. 
As an application of our findings at generic $X$ and $V$, here we report in Fig. \ref{fig1} the phase diagram obtained at $V=0$, where $K$ is given by (\ref{Kex}). In this case, the comparison with the results obtained  within the standard weak coupling approach of section 3 shows the appearance of a further non-trivial HI phase for $\frac{4\pi}{3}X\leq U_0\leq\frac{16}{\pi} X$. From Table \ref{table1} it is seen that the latter corresponds to a non-vanishing string order parameter in the charge sector, which is known to amount to the presence of degenerate edge modes \cite{MDIR}.

 \section{Conclusions}
We have derived the bosonization phase diagram of a unit density balanced two-component Fermi gas with correlated hopping,  on-site, and long-range dipolar interactions. The power law decay of  the dipolar interaction with exponent greater than 1 allows to re-sum its contributions into an effective short range potential. Moreover, spin-charge coupling terms can be neglected having, in general, larger scaling dimension. The resulting bosonized Hamiltonian consists of two spin-charge separated sine Gordon models, which phase diagram can be discussed according to Table \ref{table1}, depending on mass and Luttinger parameters. We derived the sine-Gordon models both within standard bosonization, in which case the single component Hamiltonians were the non-interacting up and down spin models; and by including non-perturbatively part of the interaction already at the level of the single component Hamiltonians. In the latter case further features emerge in the phase diagram, noticeably a non-trivial Haldane charge gapped phase also in absence of dipolar interaction. We expect that other 3- and 4-body processes \cite{DM} could be included  non-perturbatively within the LL regime of the single component Hamiltonian, possibly inducing further orders in the ground state phase diagram.
The present results should be compared with numerical findings. For instance in \cite{FMRB} by means of a density matrix renormalization group \cite{white} analysis  it was found that further exotic phases, not present in the classification given here (Table \ref{table1}), appear. In such cases, one should go beyond one loop bosonization, including higher order harmonics which were neglected here. Finally it is relevant to underline that the previous quantum phases could be studied and probed \cite{endress} with the currently available experimental setups  as proposed in \cite{FMRB}.

\end{document}